# A Preliminary Survey of Knowledge Discovery on Smartphone Applications (apps): Principles, Techniques and Research Directions for E-health

Wu-Chen Su

*National Cheng Kung University*
*Tainan, Taiwan*
*email: wuchen_sue@hotmail.com*

*Abstract*—People usually seek out varied information to deal with their health problems. However, the large volume of information available may present challenges for the public to distinguish good from suboptimal advice. How to ensure the right information for the right person at the right time and place has always been a challenge. For example, smart phone application vendor markets provide a varied selection of health applications for users. However, there is a lack of substantive reference information for consumers to base well-informed decisions about whether or not to adopt the applications they review and to ascertain the validity of the information provided by these e-health solutions. Thus, this study aims to review the existing relevant research about smart phone applications and identify pertinent research questions in the field of knowledge discovery for health applications that can be addressed in future research. Therefore, this study can be seen as an important step for researchers to explore this domain and extend our work for the well-being of public.

*Index Terms - Review; Health; Mobile Application; Knowledge Discovery; Social Network Analysis*

## I. INTRODUCTION

A rapid development of smart phone applications (hereafter referred to as "apps" per the popular usage and for brevity) has brought many business opportunities for their suppliers and relevant stakeholders. However, this rapid development also causes many potential problems for their users. For example, for health apps, how to guarantee the quality of these products particularly as serious concerns regarding the safety and validity of the information offered users thereby have arisen in recent years [1-2]. Recently, there are several principles under discussion to ascertain the quality of medical mobile apps for the public, for instance, regulations drafted by U.S. FDA [3] and selected principles from HONcode [4] by TL [2]. However, none of these relevant principles are applied to regulation of the smart phone applications market for the benefits of consumer health. In addition, the large volume of information may present challenges for the public to distinguish good from suboptimal advice. Therefore, we can conclude that how to ensure the right health apps for the right person at the right time and place has always been a challenge [1-2,5-6]. This study aims to review possible techniques, applications and principles which can be used for clustering, validating and classifying health smart phone apps across vendor markets and discuss how to develop relevant studies that facilitate consumers' search and selection of e-health apps thus contributing to the greater public health in the near future. In addition, this study is a first study to explore knowledge discovery via health apps which provides many unique research insights from different perspectives (e.g., social network analysis approach) to broaden the scope of this line of research.

## II. RESEARCH METHOD

Current research of health apps are focusing on manually analyzing small scale health apps to find evidence of their effectiveness for human beings. In order to do further investigation, we first analyzed and compared these studies to see if there are any further possibilities to do large scale analyses (e.g., classification and clustering) in future e-health research.

### A. Identifying Relevant Studies

For this study, IS and CS studies were searched to find any manual and (semi-) automatic analysis approach for health apps. In addition, a pilot review of current analytic results involving medical apps research is also provided. In May, 2013, data was collected from five major scientific databases—*IEEE Xplore, ScienceDirect, ACM Digital library, SpringerLink and PubMed.*

### B. Selection Criteria

1. In this search the keywords selected were "review", "apps", "health" and ("Android" or "Apple").

2. In addition, studies of the use of existing e-health solutions, design of new apps for particular experimental treatments, studies not in the English language, studies whose app selection criteria are not clear and those studies not providing full-text articles via host organization's online library system were all excluded. Our aim is to find research focusing on a "review" of health apps on major public app vendor markets [7].

Two evaluations of the relevant research literature were conducted. In the first run, the author screened the titles and abstracts of all relevant articles found to select possible studies. Selections were made with the expressed agreement of the authors of those studies and collaborative experts. After that, a full-text study of all possible candidate cases was made to determine the final selection.

## III. RESULT

After reviewing these articles and filtering out the studies which did not meet the aforementioned criteria, 30 articles on pertinent to this study remain. The results of this analysis are listed and discussed in Table I hereafter:

Table I. Published surveys of health apps

| Reference | Purpose | Target Group | Vendor | Evaluated (Found) |
|---|---|---|---|---|
| Demidowich, Lu et al. [8] | Usability rating, cost, functions and downloads (E) | Self-management of diabetes apps | A | 42(80) |
| Handel [9] | Health apps(E,R) | Varied | I,A,B | 32 |
| Anoop Rao [10] | Price, reports, wireless and sync with PC functions, ease of use and time for completing tasks (E) | Diabetes apps (self-monitoring of blood glucose data) | I | 12 |
| Kranz, Möller et al. [11] | Health and fitness apps for training support (E) | Health and Fitness apps | A | 16 |
| Mosa, Yoo et al. [7] | Features of smart phone apps from academic research (E) | Varied (depending on subjects) | All | 83 |
| Rusin, Årsand et al. [12] | Relevant nutrition functions of apps and web applications (E) | The general population and people with diabetes | I,A | 45 (iOS:26 and Android:19) |
| Deveau and Chilukuri [13] | Dermatology apps (E,R) | Varied (patients, dermatologists and medical students) | I,A | 16 |
| Verhagen, van Mechelen et al. [14] | Sport injury prevention and rehabilitation apps (E) | Varied | I | 325 |
| Rosser and Eccleston [15] | Pain-related smart phone apps (E) | Varied | All | 104(111) |
| Kharrazi, Chisholm et al. [16] | Stand-alone mobile personal health record (mPHR) apps (E) | Varied | I,B,A | 19(1771) (iOS:8, BlackBerry:5 and Android:6) |
| Székely, Talanow et al. [17] | Diagnostic imaging apps (E) | Varied (diagnostic imaging professionals, radiographers and residents and people in this field) | I,A,B,W | 81(102) |
| Gomez-Iturriaga, Bilbao et al. [18] | Radiation oncology and oncology apps (E) | Varied | I,A | N/A |
| Chomutare T [19] | Diabetes care apps (E) | Varied | I,A,B,N | 137(973) |
| Martínez-Pérez B [20] | Apps about WHO defined diseases[3] (E) | Varied | All | 16(3673) |
| Muessig KE [21] | HIV/STD-related apps (E) | Varied | I,A | 55(1937) |
| West JH et al. [22] | Paid health and fitness apps (E) | Varied | I | 3336(5430) |
| Stark [23] | Cell Biology apps (E,R) | E-learning for students | A | 10(12) |
| Connor et al. [24] | Hernias apps (E) | Varied (clinicians and their patients) | All | 26 |
| Stevens et al. [25] | Weight loss surgery apps (E) | Varied (clinicians and their patients) | I,A,B | 28(38) |
| Dayer et al. [26] | Medication adherence apps (E) | Varied (pharmacists and non-adherent patients) | I,A,B | 160 |
| Carter et al. [27] | Vascular diseases apps (E) | Varied (vascular health care workers and patients) | All | 49 |
| Pandey et al. [28] | Cancer and oncology apps (E) | Varied (clinicians and their patients) | I | 77(93) |
| O'Neill et al. [29] | Colorectal disease themed apps (E) | Varied (clinicians and their patients) | All | 63(68) |
| Sondhi et al. [30] | Pediatric health care apps (E,R) | Varied | All | 43 |
| Liu et al. [31] | Medical, health and fitness apps (E,R) | Varied | I | 6(200) |
| Abroms et al. [32] | Smoking cessation apps (E) | Varied | I | 47(62) |
| Al-Hadithy et al. [33] | Education, telemedicine and global health apps (E) | Plastic surgeons | I,A | 18 |
| Boulos et al. [34] | GPS and geosocial apps (E) | Varied (children and adolescents) | All | 30 |
| Connor et al. [35] | Bariatric surgery apps (E) | Varied | All | 83(674) |
| Huckvale et al. [36] | Asthma self-management apps (E) | Varied (people with asthma, healthcare professionals and policy-makers) | All | 103(207) |

1. Vendor markets: Apple iPhone/iPad(I), Android(A), BlackBerry(B), Nokia/Symbian(N) and Windows Mobile(W)
2. E: Evaluate, R: Recommend
3. Eight of the most prevalent diseases listed by the WHO are [51]: iron-deficiency anemia, hearing loss, migraine, low vision, asthma, diabetes mellitus, osteoarthritis (OA), and unipolar depressive disorders

## A. Findings

As can be seen in the survey of the approaches listed in Table I, 8 out the 30 articles selected focus on evaluating diabetes diagnostic-related applications. The second most discussed topic is that dealing with applications offering advice on physical activities, for example, e-coach [11], injury and pain management solutions [14-15,22]. This reflects a need of our community for advice on personal health challenges. According to statistics cited in *AppBrain* [37], there are more than 28,000 health apps on the Android market. Thus, the result of this analysis represents only a very small proportion of the health apps on the market. In sum, we acknowledge that there should be some methods which can be used to help researchers to discover useful information regarding these apps, and improve the speed of research progression. To that end, researchers can put more focuses on the content of their studies which benefit society. Thus, hereafter, current proposed principles for categorizing medical smart phone apps and reviewing techniques for recommendations of apps and methods of knowledge discovery concerning the smart phone app market are introduced to help us understand the gap between the information available to app users and suppliers.

## IV. DISCUSSION

### A. Recommendation of Apps

Users tend to use ways provided by vendors to find desirable apps. However, inexperienced users using less than optimal search methods and/or inadequate descriptions of apps may lead to unexpected results and incur associated cost. Thus, several studies endeavor to deal with this issue. The research by Jiang, Vosecky et al. [38] provides a semantic-based search and ranking method for Android vendors to resolve problems of poorly organized descriptions of apps and poor ranking results, also claiming that similar circumstances exist among other major vendors. Instead of recommending new apps, Yin, Luo et al. [39] consider the needs of replacing an old app with a new one. They used an estimated tempting value of a new app and compared that with actual satisfactory value of the old app recommended for replacement. Lulu and Kuflik [40] propose an unsupervised learning algorithm for eliciting descriptions of apps and information from professional blog sites on the Internet to cluster apps and demonstrate them hierarchically for ease of search. Böhmer, Ganev et al. [41] argue that the performance of a recommendation system should be evaluated through the life cycle of a user's engagement with a mobile app, reviewing not only the acceptance of new apps by users, but also the stages of viewing, installation, direct usage and long-term usage to get insight into the reactions of user in each step. In addition to previous works, a hybrid social recommendation system which employs tag and context information is purposed in [42], whose authors designed an app for users to use in background mode interactively with a recommendation system on the Internet to achieve ideal results and deal with first-rater, cold-start and sparsity problems. However, none of the research analyzed focuses on recommendation of health apps. More specifically, users can only consider the clustering and classification problems of apps according to varied principles (e.g., FDA[3]) first. After clustering and classification of apps, a user can then effectively search among apps manually or adopt suggestions from any recommendation system based on the organized hierarchy. In addition, the explicit information given for apps, for instance, descriptions and comments are not enough to be used in the recommendation stage. The major considerations are correctness of relevant health information and potential malicious behavior of these apps. Further investigation before recommending health apps for users is necessary.

### B. Knowledge Discovery on the Android Market

In order to understand the apps marketed by the major vendors (e.g., Android Google Play and Apple App Store), there are ever more studies starting to analyse the information provided by, and different features of, apps. The Android system is designed with open source concepts and becoming the first choice for conducting research and building applications among academia and industry. Thus, our research focuses on relevant analytic methods and technical analyses of apps on the Android market.

#### 1) Security and Privacy Concerns of Apps

To date, a massive amount of knowledge (and relevant tools) is discussed, as provided by interested people, directly or indirectly helping consumers understand these apps from different perspectives. For example, app developers can use open-source APIs to download apps and their relevant information from the Android market [43]. After that, they can reverse-engineer apps into Java source codes [44]. Thus, there are varied studies using these tools to evaluate and help people to understand the nature of these apps, such as a study of different privacy behaviours of apps conducted to help people understand their impact on users. More specifically, for further analysis, researchers have decomposed the codes by which apps are written and extracted their APIs as the designers have used and linked them to privacy-behaviours of users' activities. In their experimnets, they conducts a static analysis on nearly 80,000 apps available from Android vendors, sent the results to users and asked for their feedback of any potential effect on their adoption patterns [45].

In addition, security concerns regarding use of apps is also an important topic to be explored and discussed by researchers. The current major app vendor markets do not provide ideal verification mechanisms to check for malicious behaviour of downloaded apps. Thus, an analysis was conducted on 47 SMS-based Android apps in regard to their potenial to be used for theft of users financially related data and detected there are nearly 90% of apps that have the potential for such issues [46]. In addition, Zhou et al. [47] find that some apps are designed for stealing advertisement revenues from the app designers and to allow for remote control of victims' smart phones.

#### 2) Network Concepts of Apps

In addition, the potential for leak of information allowed by some suspicious apps has been discovered and discussed by their destinations (networks) from around 4,000 apps on the Android market. According to the report by Rastogi, Chen et al. [48], the designers of these apps and information flow from these apps to the problematic domains on the Internet can be seen as network structures. Multiple vulnerable apps use these domains as the destination that provides sensitive information (e.g., GPS information, contacts and IMEI/IMSI number) of their users for unknown purposes. Those researchers conclude that the authors of these free apps use third-party advertisement libraries which in turn causes their app to fall victim to malicious behavior designed to leak the personal private information of their users.

### 3) Dynamic Analysis

In addition to previous work, Zheng, Zhu et al. [49] focus on dynamic flow analyses of programs, claiming the static analysis method cannot effectively detect full program behaviours. The authors design a tool for analysing UI (User Interface) interactions behind these apps and effectively detected paths which lead to sensitive information. To sum up, we conclude that security and privacy issues are certainly manifest among these apps on the Android market. However, no specific research focuses on health apps. Thus, a further and complete analysis according to current accepted verification principles is necessary.

### 4) Categorization Principles of E-health Apps

In order to further categorize these e-health apps, we have to understand current developments of relevant principles on the market. These potential principles are designed by abstract concepts of consumer usage, target groups, subject content of apps or disease-based perspectives intended to protect consumer health. Hereafter listed are details of aims, intended end users and principles for each categorization approach in Table II. Obviously, one cannot find that any major vendor or relevant health informatics research has applied these features to (semi-) automatic selections of qualified medical apps for providing better user experiences. Thus, we argue a first academic approach for validating these medical apps should be purposed and tested for the benefits of public health.

Table II. Published principles for categorizing e-health apps

| Reference | Aims | End user |
|---|---|---|
| FDA [3] | U.S. FDA intends to apply below principles for its authority to different types of health mobile apps | Industry and Food and Drug Administration Staff |
| | **Principles** (a) Mobile medical apps that are extensions of regulated medical device for purposes of controlling the medical device or for the purpose of displaying, storing, analyzing, or transmitting patient-specific medical device data. (b) Mobile medical apps that transform or make the mobile platform into a regulated medical device by using attachments or sensors or similar medical device functions. (c) Mobile medical apps that allow the user to input patient-specific information and - using formulae or a processing algorithm - output a patient-specific result, diagnosis, or treatment recommendation that is used in clinical practice or to assist in making clinical decisions. | |
| TL [2] | Building a systematic self-certification model for mobile medical apps | Health care professional and app designers |
| | **Principles** Information must be authoritative, Purpose of the website (to vendor market), Confidentiality, Information must be documented: referenced and dated, Justification of claims, Contact details, Financial disclosure and Advertising policy. | |

## V. CONCLUSIONS AND FUTURE RESEARCH DISCUSSION

### A. Research Direction of Profiling Health Apps

Recent research [1,5-6] has discussed several suggestions to resolve health apps overload and quality issues. For example, management policies (e.g., standardized medical information and use of authorized open source medical data for app development) and peer review of the content of health apps. However, these potential measurements or movements could take time and large resources to be effective. Based on this current research's review, these current studies provide us deep insight into manual analyses of health apps, principles for categorizing medical apps and methods for uncovering issues of concern as well as management of security and privacy issues of products on the Android market. Even so, these studies cannot be used as a certification framework for health apps. As aforementioned, different principles and concerns may have different methods of clustering and classification of apps. From the author's perspective, some possibilities should be sought out to address these questions effectively.

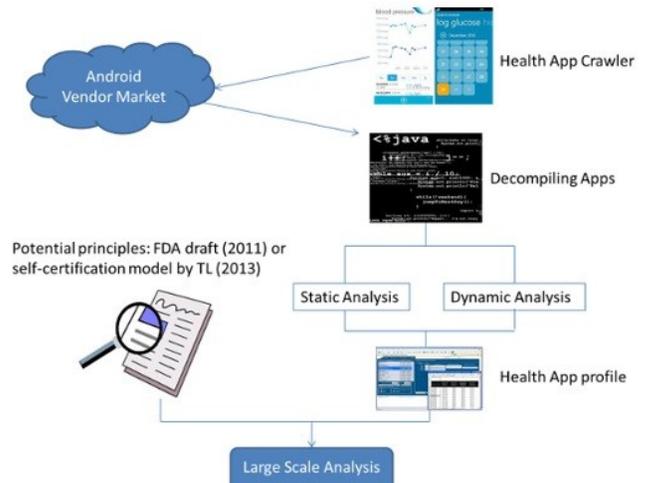

Fig. 1. A systematic research flow

For example, in Fig. 1, an app crawler is designed to obtain e-health apps from the Android market and decompile codes of these apps by open source tools. Secondly, newly developed algorithms for supporting statistic and dynamic analyses of app content will be proposed to understand the profiles of health apps on the Android market. Based on the creation of app profiles and the current potential principles for categorizing medical apps, a further large scale analysis will be executed to cluster, validate and classify e-health apps on the Android market.

### B. Research Direction of Analysing Social Network Patterns among App Vendor Markets

In addition, given the prior discussion of network concepts of apps in the previous discussion section, there are also potential opportunities for analyzing these vendor markets by utilizing social network analysis approaches [50]. In Fig. 2 and Fig. 3, one can clearly see the relationship between each actor in the network according to their roles. These relationships can be seen as explicit and/or implicit networks. In Fig. 2, an app creator can design two different versions of similar functioning apps and market them both via Google's Play and Apple's App Store. In addition, they can also cooperate with advertising companies and put their advertisements in their apps to gain extra benefits. Then, users can download, use and/or evaluate these apps if they want to do it. If these apps have malicious designs, they may detect a user's private information and send it back to these advertisement networks for unknown purposes. Furthermore, as seen in Fig. 3, designer 1 and designer 2 are working in the same organization which means they may share their knowledge and experience of designing apps. Considering these relationships, they can learn from each other and apply new skills for future designs to attract more supported users and produce more creative apps. Thus, a deep analysis can help us to utilize this powerful information to deal with issues of security and privacy among all these apps and vendor markets. Future research may arise from these scenarios. For example,

1. *How malicious (beneficial) networks are structured (clustered) currently or being developed in the future (perdition) among e-health apps (e.g., medical, fitness, etc.), app designers, advertisement networks and apps users?*

Further studies are required for exploring this theme for the benefits of consumer health.

In conclusion, this review has reviewed current methods, approaches (manual and automatic) and principles which can be used for clustering, validating and classifying health smart phone apps. In addition, it is here suggested that a social network analysis approach would help researchers to address the positive and negative impact on e-health apps from different perspectives and allow them to take necessary actions for avoiding or even contrasting these mean behaviors of apps across these popular markets for promoting the level of reliable and comfortable user experience. Most importantly, support for well-informed, intelligent decision by users in selecting appropriate e-health apps is possible in the future.

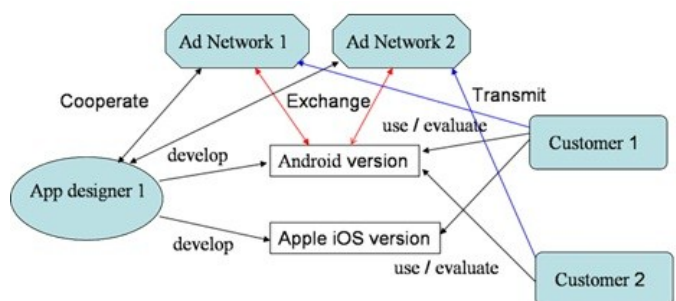

Fig. 2. Example of relation networks between multiple actors and apps

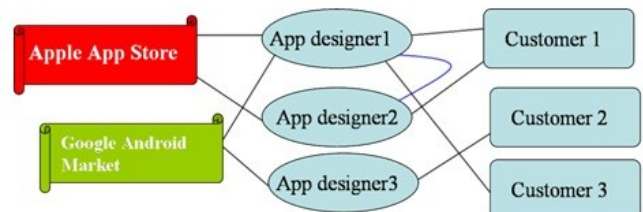

Fig. 3. Example of relation networks across of multiple vendor markets